\documentclass[twocolumn,longauthor]{aastex62}

\newcommand{\omc}{OMC-1}
\newcommand{\Blos}{\ensuremath{B_{\text{los}}}}
\newcommand{\Bpre}{\ensuremath{B_{{\rm pre}}}}
\newcommand{\Bpost}{\ensuremath{B_{{\rm post}}}}
\newcommand{\meth}{CH$_3$OH}
\newcommand{\jybeam}{Jy~beam$^{-1}$}
\newcommand{\kmS}{km~s$^{-1}$}

\shorttitle{25 GHz Methanol Maser Zeeman Effect toward \omc}
\shortauthors{Sarma \& Momjian}

\begin{document}

\title{\textsc{A Curious Case of Circular Polarization in the 25 GHz \\ Methanol Maser Line toward OMC-1}}

\author{A.~P.~Sarma}
\affiliation{Department of Physics and Astrophysics, DePaul University, \\ 2219 N.~Kenmore Ave., Byrne Hall 211, Chicago, IL 60614; asarma@depaul.edu}

\author{E.~Momjian}
\affiliation{National Radio Astronomy Observatory, P.O~Box O, Socorro, NM 87801}

\accepted{December 12, 2019 for publication in The Astrophysical Journal}

\begin{abstract}
We report the detection of a circular polarization signature in the Stokes~$V$ profile of a 25 GHz Class~I \meth\ maser toward the high mass star forming region \omc. Such a feature usually constitutes a detection of the Zeeman effect. If due to a magnetic field in \omc, this would represent the first detection and discovery of the Zeeman effect in the 25~GHz Class I \meth\ maser. The feature in Stokes~$V$ is detected in two observations with different angular resolutions taken eight years apart with the Very Large Array (VLA); for our 2009 D-configuration observations, the fitted value for $z\Blos$  is $152\pm12$~Hz, where $z$ is the Zeeman splitting factor and \Blos\ is the line-of-sight magnetic field. For our 2017 C-configuration observations, the fitted value for $z\Blos=149\pm19$~Hz, likely for the same maser spot. These correspond to \Blos\ in the range 171-214~mG, depending on which hyperfine transition is responsible for the maser line. While these \Blos\ values are high, they are not implausible. If the magnetic field increases in proportion to the molecular hydrogen density in shocked regions, then our detected fields predict values for the pre-shock magnetic field that are in agreement with observations. With \Blos=171-214~mG, the magnetic energy in the post-shocked regions where these 25~GHz Class~I \meth\ masers occur would dominate over the kinetic energy density and be at least of the order of the pressure in the shock, implying that the magnetic field would exert significant influence over the dynamics of these regions.

\end{abstract}


\section{Introduction}

\begin{deluxetable*}{lcccccrrrrcrl}
\tablenum{1}
\tablewidth{0pt}
\tablecaption{PARAMETERS FOR VLA OBSERVATIONS
	\protect\label{tOP}}
\tablehead{
& \colhead{2009 Observations} & \colhead{2017 Observations} \\
\colhead{Parameter} & 
\colhead{Value} & \colhead{Value}}
\startdata
Date &  2009 Oct 22 & 2017 Jun 11  \\
Configuration & D & C \\
R.A.~of field center (J2000) & 05$^{\rm h}$ 35$^{\rm m}$ 14$\fs$02 & 05$^{\rm h}$ 35$^{\rm h}$ 14$\fs$02 \\
Dec.~of field center (J2000) & $-5\arcdeg$ 22$\arcmin$ 30$\farcs$9 & $-$5$\arcdeg$ 22$\arcmin$ 30$\farcs$9 \\
Total bandwidth (MHz) & 1.56 & 4.00 \\
No.~of channels & 255 & 1024 \\
Channel spacing (km~s$^{-1}$) & 0.073 & 0.094\,\tablenotemark{a} \\
Approx.~time on source (hr) & 2.4 & 2.3 \\
Rest frequency (MHz) & 24959.079 & 24959.079 \\
FWHM of synthesized beam & 3$\farcs$34 $\times$ 2$\farcs$71 
& 1$\farcs$30 $\times$ 0$\farcs$89 \\
& P.A. = $-$2$\fdg$61 & P.A. = 19$\fdg$47 \\
Line rms noise (mJy~beam$^{-1}$)\,\tablenotemark{b} & 4.0 & 2.2  \\ 
\enddata
\tablenotetext{a}{Image cubes were made by averaging every two channels.} 
\tablenotetext{b}{The line rms noise was measured from the Stokes $I$ image cube using maser line free channels.}
\end{deluxetable*}

Of the well known Class I methanol (\meth) maser transitions, the $J_2 - J_1$~E series of maser lines near 25 GHz are different because of their inversion mechanism. Unlike Class I \meth\ maser transitions at  36, 44, 84, and 95 GHz, the $J_{k=2} - J_{k=1}$ lines responsible for \meth\ maser transitions near 25 GHz depend on $\Delta k \ne 0$ to build up the population in the $k$=2 and $k$=1~E-type methanol ladders \citep*{leurini+2016}. Like all the other Class I \meth\ masers, however, masers near 25 GHz also arise in outflows in star forming regions, where collisional pumping creates the necessary population inversion\ \citep{menten1993}. Therefore, they enable us to observe such regions at high angular resolution. In particular, high mass star forming regions must be observed at high angular resolution because high mass stars usually form in clusters, and much remains to be known about how they form\ \citep{motte+2018}. Moreover, both 36 GHz and 44 GHz Class I \meth\ masers trace the Zeeman effect\ \citep{sarma+2009, momjian+2019}, which constitutes the most direct method to measure magnetic fields in star forming regions\ \citep[e.g.,][]{troland+2008}. The ability to detect the Zeeman effect in 25 GHz Class I \meth\ maser lines would open a new window into the measurement of magnetic fields in star forming regions. Magnetic fields are known to play an important role during several stages of the star formation process, but the details remain a matter of debate, in part due to the scarcity of observational data in the high density environments close~to~the~forming star\ \citep{crutcher2012, krumholz+2019}.

The Orion Molecular Cloud 1 (\omc) holds the distinction of hosting, to date, the brightest masers in the 25 GHz Class I \meth\ species\ \citep{barrett+1971}. \omc\ is a ridge of dense molecular gas that lies behind the ionized region caused by the OB stars of the Trapezium cluster in the Orion Nebula \citep[e.g.,][]{pabst+2019}. With an extent of $\sim$2~pc along the north-south direction, \omc\ is near the center of the integral-shaped filament that was mapped in $^{13}$CO by \citet{bally+1987}. This integral-shaped filament, at the northern end of the Orion A giant molecular cloud, is a clumpy, narrow ($<1\arcmin$, or 0.2~pc) ridge of emission that extends in the north-south direction for over $50\arcmin$ (7~pc), with fainter filaments and clumps extending orthogonal to the ridge for several arcmin\ \citep{johnstone+1999}. Being part of the nearest high mass star forming region at a distance of $388\pm5$~pc (\citealt{kounkel+2017}), \omc\ has been observed extensively at optical, infrared, radio, etc., wavelengths in both continuum and spectral lines\ \citep[see, e.g.,][and references therein]{genzel+1989, bally2008, hacar+2018}. \omc\ contains the well-known Becklin-Neugebauer (BN) object\ \citep{becklin+1967} and the Kleinmann-Low (KL) nebula\ \citep{kleinmann+1967}. Centered near Orion-KL is a wide-angle, high-velocity outflow oriented in a northwest-southeast direction\ \citep{erickson+1982}. When this high velocity outflow slams into ambient gas, it produces shocks; such shocked regions can be imaged in the molecular hydrogen $\nu=1-0$~S(1) emission line, as has been done for \omc\ \citep{beckwith+1978}. It is in such shocked regions that \meth\ masers are formed; in \omc, the 25~GHz Class~I \meth\ masers are distributed along an arc that runs from northwest to southeast\ \citep{johnston+1992}. 

In this paper, we report the presence of a circular polarization signature in the Stokes $V$ profile of a 25 GHz Class I \meth\ maser line toward \omc\ that is usually interpreted as a detection of the Zeeman effect. If this Stokes $V$ profile is truly due to a magnetic field, then the work reported in this paper would represent the first detection of the Zeeman effect in the 25 GHz Class I \meth\ maser line. In Section \ref{sec:obs}, we describe the observational setup and the data reduction process. In Section \ref{sec:res}, we present our results, along with a description of the analysis of data for the Zeeman effect. These results are discussed in Section \ref{sec:disc}, and our conclusions are presented in Section \ref{sec:conc}.

\section{Observations and Data Reduction} 
\label{sec:obs}

We present results from two separate epochs of observations on the 25~GHz ($5_2-5_1$~E) Class I \meth\ maser line (rest frequency 24.959~GHz) toward the star forming region \omc. Observations in the first epoch were carried out with the pre-upgrade Very Large Array (VLA)\footnote{The National Radio Astronomy Observatory (NRAO) is a facility of the National Science Foundation operated under cooperative agreement by Associated Universities, Inc.} on 2009 October 22 in a single 3\,hr session in the D-configuration (maximum baseline of $\sim$1~km). In 2009, the VLA was still undergoing the major upgrade through the Expanded VLA (EVLA) project. While the observations used all the 21 antennas retrofitted to the EVLA standards and excluded the old style antennas, the data were correlated using the old VLA correlator delivering a bandwidth of 1.56 MHz with 255 spectral channels and dual polarization products (RR, LL). This resulted in a channel spacing of 6.1 kHz, which corresponds to 0.073~\kmS\ at the observed frequency. At the time, the use of EVLA antennas with the old correlator in spectral line observations introduced aliasing that impacted the lower 0.5~MHz of the bandwidth. Therefore, the 25~GHz \meth\ maser line was centered in the upper half of the 1.56~MHz bandwidth. Observations in the second epoch were carried out with the post-upgrade VLA, formally the Karl G. Jansky Very Large Array, on 2017 June 11, also in a single 3~hr session but in the more extended C-configuration (maximum baseline of 3.4~km). The Wideband Interferometric Digital ARchitecture (WIDAR) correlator, which is the post-upgrade VLA correlator, was configured to deliver a single 4~MHz sub-band with dual polarization products (RR, LL) and 1024 spectral channels. The resulting channel spacing was 3.91~kHz, corresponding to 0.047~\kmS\ at the observed frequency. In both epochs, the source J0542+4951 (3C147) was observed to calibrate the absolute flux density scale. In the higher angular resolution observations of 2017, we employed phase referencing in part of the observing session in order to derive the absolute positions of the masers in the target source. The phase calibrator was the nearby source J0541$-$0541, and the phase referencing cycle time was 5~min. All the data reduction steps, calibration, imaging and deconvolution, were carried out independently for each observing epoch using the Astronomical Image Processing System (AIPS; \citealt{greisen+2003}) of the NRAO. The spectral channel with the brightest maser emission was split off, and self-calibrated first in phase, then in both phase and amplitude, and imaged in a succession of iterative cycles. The final self-calibration solutions were then applied to the full spectral-line uv data sets of \omc\ from each epoch. In order to improve the signal-to-noise of the higher angular resolution data set of 2017, the Stokes $I$ and $V$ image cubes were constructed by averaging every two channels, resulting in a velocity spacing of 0.094\,\kmS.  Note that AIPS calculates the Stokes parameter $I$ as the average of the right circular polarization (RCP) and left circular polarization (LCP), so that $I=$~(RCP + LCP)/2, whereas Stokes $V$ is calculated by AIPS as half the difference between RCP and LCP, so that $V=$~(RCP~$-$~LCP)/2; henceforth, all values of $I$ and $V$ are based on this implementation in AIPS. Also note that RCP is defined here in the standard radio convention, in which it is the clockwise rotation of the electric vector when viewed along the direction of wave propagation. Table~\ref{tOP} summarizes the parameters of the VLA observations and the corresponding synthesized beamwidths and other parameters for each observing epoch.

\begin{figure}[b!]
\plotone{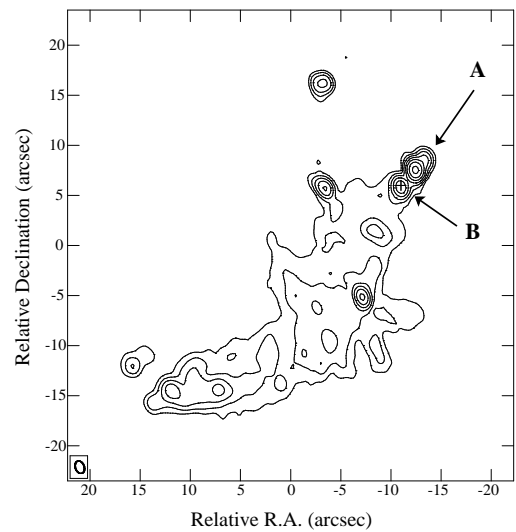}
\caption{Velocity-integrated image of the 25 GHz Class I \meth\ masers toward \omc\ from our 2017 VLA C-configuration observations. The contours are at $(0.5, 1, 2, 4, 8, 16, 32) \times 0.4$~\jybeam~\kmS. The FWHM of our synthesized beam is 1$\farcs$30 $\times$ 0$\farcs$89, and is shown as an inset in the bottom left. The (0,0) position in this image corresponds to $\alpha = 05^{\rm h}~35^{\rm m}~14\fs455, \delta = -05\arcdeg~22\arcmin~31.40\arcsec$ (J2000). The parameters of masers A and B marked in this figure are given in Table~\ref{tGauss}. The plus sign marks the position in maser B toward which we measured the Zeeman effect. \label{fig1}}
\end{figure}

\section{Results}
\label{sec:res}

\begin{figure}[htb!]
\plotone{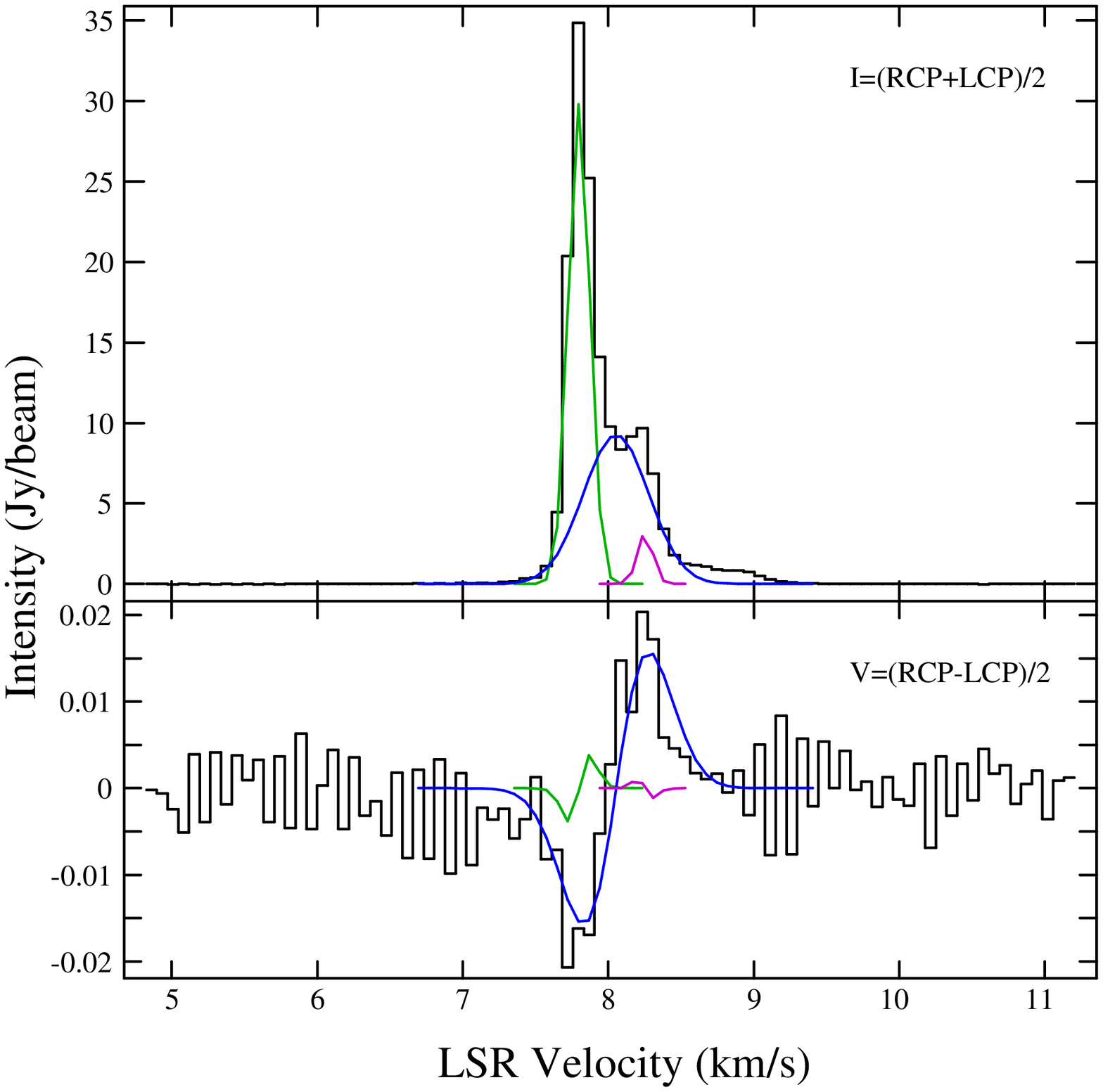}
\caption{Stokes $I$ (\textit{upper panel---black histogram-like line}) and Stokes $V$ (\textit{lower panel---black histogram-like line}) profiles from our 2009 observations toward the maser spot in \omc\ listed as A+B in Table \ref{tGauss}. The green, blue, and magenta curves in the upper panel show the Gaussian components that we fitted to the Stokes $I$ profile (components 1, 2, and 3, respectively, for the maser listed as A+B in Table \ref{tGauss}). The green, blue, and magenta curves in the lower panel are the derivatives of the corresponding colored curves in the upper panel, scaled by the fitted value of $z\Blos$ for each curve, obtained from our fitting procedure  described in Section~\ref{sec:res}. \label{fig2}}
\end{figure}

\begin{figure}[htb!]
\plotone{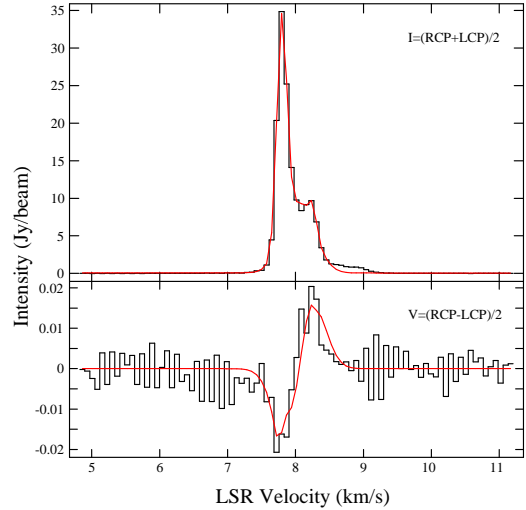}
\caption{Stokes $I$ (\textit{upper panel---black histogram-like line}) and Stokes $V$ (\textit{lower panel---black histogram-like line}) profiles from our 2009 observations toward the maser spot in \omc\ listed as A+B in Table \ref{tGauss}. The red curve in the upper panel is the sum of the three Gaussian components shown by green, blue, and magenta curves in the upper panel of Figure~\ref{fig2} (and listed in Table~\ref{tGauss}) that we fitted to the Stokes $I$ profile. The red curve superposed on the Stokes $V$ profile in the lower panel is the sum of the green, blue, and magenta curves shown in the lower panel of Figure~\ref{fig2}; that is, it is the sum of the scaled derivatives of the Gaussian components fitted to the Stokes $I$ profile, where each of the three derivative profiles has been scaled appropriately by the fitted value of $z\Blos$, as described in the caption to Figure~\ref{fig2}. \label{fig3}}
\end{figure}

The 25~GHz Class I \meth\ masers toward \omc\ are arranged in an arc that spans roughly a quarter of a circle and runs from northwest to southeast; this is shown in Figure~\ref{fig1}, and corresponds well with \citet{johnston+1992}. Figure~\ref{fig1}\ is a velocity-integrated image taken from our VLA C-configuration observations; in our D-configuration observations of 2009, the maser marked with a plus ($+$) sign and the maser to its northwest (masers B and A respectively in Figure~\ref{fig1}) were observed as one (unresolved) maser feature. The upper panel of Figure~\ref{fig2}\ shows the Stokes $I$ profile toward this maser from our D-configuration observations in 2009. We fitted this profile with three Gaussian components; they are also displayed in the upper panel of Figure~\ref{fig2}. The intensity, velocity at line center with respect to the LSR, and velocity linewidth measured at full width at half maximum (FWHM) of each of the three components are given in Table~\ref{tGauss}, where we've listed the maser spot as A+B in view of our higher resolution observations carried out in 2017 (and explained in more detail below). Of the three components for maser~A+B in our 2009 observations, the strongest (29.8~\jybeam) is centered at an LSR velocity of 7.80~\kmS, component 2 (9.3~\jybeam) is centered at 8.05~\kmS, and component 3 (3.1~\jybeam) is centered at 8.25~\kmS. Components 1 and 3 are quite narrow with FWHM velocity linewidths of 0.17~\kmS\ and 0.13~\kmS\ respectively, whereas component 2 is broader with a FWHM linewidth of 0.53~\kmS. The composite profile obtained from the sum of these three Gaussian components is shown in the upper panel of Figure~\ref{fig3}. 

\begin{deluxetable*}{cccccrrrrrcrl}
\tablenum{2}
\tablewidth{0pt}
\tablecaption{FITTED PARAMETERS FOR \omc\ MASERS
	\protect\label{tGauss}}
\tablehead{
\colhead{Maser} &
\colhead{Component} &
\colhead{Intensity} & 
\colhead{Center Velocity\tablenotemark{a}} &
\colhead{Velocity Linewidth\tablenotemark{b}} \\
\colhead{} &
\colhead{} &
\colhead{(Jy~beam$^{-1}$)}  &
\colhead{(km~s$^{-1}$)} &
\colhead{(km~s$^{-1}$)}}
\startdata
\multicolumn{5}{c}{2009 observations} \\ \hline
A+B &1 & $29.84 \pm 0.46$  & $7.799 \pm 0.001$ & $0.173 \pm 0.003$ \\ 
& \textbf{2}\tablenotemark{c} & \phantom{1}$\mathbf{9.27 \pm 0.20}$ & $\mathbf{8.054 \pm 0.013}$ & $\mathbf{0.529 \pm 0.016}$ \\ 
& 3 & \phantom{1}$3.13 \pm 0.36$ & $8.254 \pm 0.006$ & $0.127 \pm 0.019$ \\ \hline\hline 
\multicolumn{5}{c}{2017 observations} \\ \hline
A & 1 & \phantom{1}$54.30 \pm 1.30$ & $7.776 \pm 0.001$ & $0.162 \pm 0.002$ \\ 
& 2 & \phantom{1}$14.56 \pm 1.14$ & $7.862 \pm 0.008$ & $0.293 \pm 0.007 $ \\ \hline 
B & 1 & \phantom{1}$12.10\pm0.91$ & $8.210\pm0.002$ & $0.234 \pm 0.008$ \\ 
& \textbf{2}\tablenotemark{c} & \phantom{1}$\mathbf{4.47 \pm 0.61}$ & $\mathbf{8.084 \pm 0.039}$ & $\mathbf{0.487 \pm 0.033}$ \\ 
& 3 & \phantom{1}$0.66 \pm 0.15$ & $8.649 \pm 0.088$ & $0.395 \pm 0.130$ \\ 
\enddata
\tablenotetext{a}{ The center velocity values are with respect to the LSR.}
\tablenotetext{b}{ The velocity linewidth was measured at full width at half maximum (FWHM).}
\tablenotetext{c}{The components marked in bold (component 2 for maser A+B from our 2009 observations, and component 2 for maser B from our 2017 observations) are those in which we have significant detection for $b=z\Blos$.}
\end{deluxetable*}

\begin{figure}[htb!]
\epsscale{1.06}
\plotone{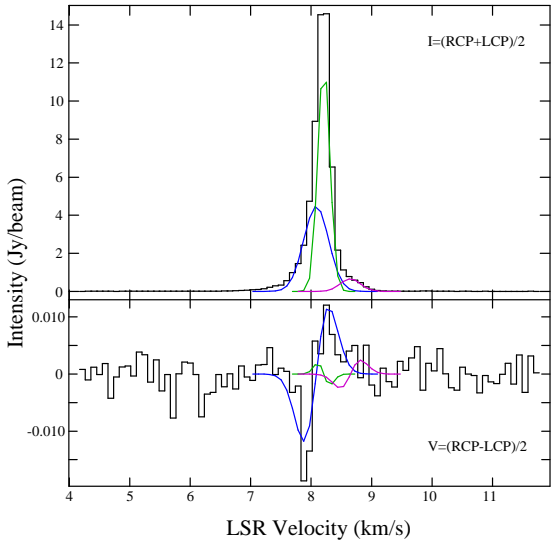}
\caption{Stokes $I$ (\textit{upper panel---black histogram-like line}) and Stokes $V$ (\textit{lower panel---black histogram-like line}) profiles from our 2017 observations toward the maser in \omc\ listed as B in Table \ref{tGauss}. The green, blue, and magenta curves in the upper panel show the Gaussian components that we fitted to the Stokes $I$ profile (components 1, 2, and 3, respectively,  for maser B in Table \ref{tGauss}). The green, blue, and magenta curves in the lower panel are the derivatives of the corresponding colored curves in the upper panel, scaled by the fitted value of $z\Blos$ for each curve, obtained from our fitting procedure  described in Section~\ref{sec:res}.  \label{fig4}}
\end{figure}
\begin{figure}[ht!]
\epsscale{0.99}
\plotone{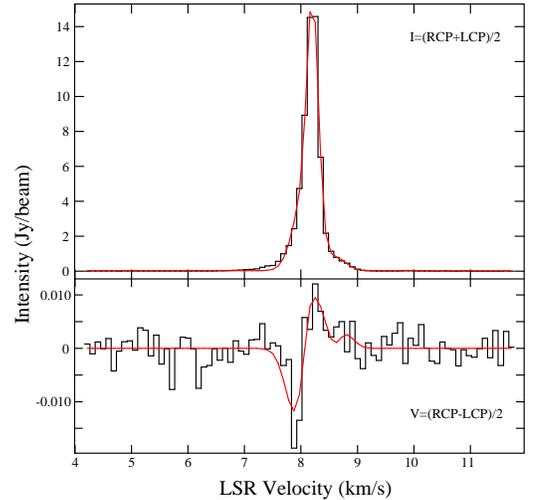}
\caption{Stokes $I$ (\textit{upper panel---black histogram-like line}) and Stokes $V$ (\textit{lower panel---black histogram-like line}) profiles from our 2017 observations toward the maser in \omc\ listed as B in Table \ref{tGauss}. The red curve in the upper panel is the sum of the green, blue, and magenta Gaussian components shown in Figure~\ref{fig4} (and listed in Table~\ref{tGauss}) that we fitted to the Stokes $I$ profile. The red curve superposed on the Stokes $V$ profile in the lower panel is the sum of the three colored curves shown in the lower panel of Figure~\ref{fig4}; that is, it is the sum of the scaled derivatives of the Gaussian components fitted to the Stokes $I$ profile, where each of the three derivative profiles has been scaled appropriately by the fitted value of $z\Blos$, as described in the caption to Figure~\ref{fig4}. \label{fig5}}
\end{figure}

In our higher angular resolution (VLA C-configuration) observations of 2017, the maser spot A+B from our 2009 observations was resolved into two distinct spots, both of which have additional velocity components; we have labeled the stronger maser spot as A and the lower intensity spot as B (see Figure~\ref{fig1}). We fitted maser A from our 2017 observations with two Gaussian components. Component 1 (54~\jybeam) is centered at LSR velocity 7.78~\kmS\ and has a velocity linewidth of 0.16~\kmS. On the basis of the velocity at line center and the FWHM linewidth, component 1 of maser A from our 2017 observations likely matches component~1~of~maser~A+B~from~our~2009~observations.~Maser~A~in our 2017 observations also has a second weaker component (14.6~\jybeam) at an LSR velocity of 7.86~\kmS\ and a FWHM velocity linewidth of 0.29~\kmS. Meanwhile, we fitted the lower intensity maser spot B from our 2017 observations with three Gaussian components (Table~\ref{tGauss}); the Stokes $I$ profile for maser B, together with these three Gaussian components, is shown in the upper panel of Figure~\ref{fig4}. The strongest of these three velocity components (12.1~\jybeam) is at an LSR velocity of 8.21~\kmS, component 2 is at 8.08~\kmS, and component 3 is at 8.65~\kmS. The composite profile obtained from the sum of these three Gaussian components is shown in the upper panel of Figure~\ref{fig5}.

The Stokes $V$ profile of maser A+B from our 2009 observations (lower panel of Figure~\ref{fig2} and Figure~\ref{fig3}) reveals an S-shaped structure that is usually taken to be a detection of the Zeeman effect (also see Discussion in \S~\ref{sec:disc} below). So does the Stokes $V$ profile from our 2017 observations (lower panel of Figure~\ref{fig4} and Figure~\ref{fig5}). No other compact maser spots in this field show any such feature. Whenever an S-shaped feature is observed in Stokes $V$, the magnetic field strength is usually determined by fitting a numerical frequency derivative of the Stokes $I$ spectrum to the Stokes $V$ spectrum; details are in, e.g., \citet{ms2017}. The Stokes $V$ profile is usually fit simultaneously to the derivative of the $I$ profile and a scaled replica of the $I$ profile itself via the equation (\citealt{th1982}; \citealt{sault1990}):
\begin{equation}
V = aI + \frac{b}{2}\, \frac{dI}{d\nu}  \label{e1}
\end{equation}
The scaled replica of the $I$ spectrum is included in the fit to account for small calibration errors in RCP versus LCP; for all results reported in this paper, $a \lesssim 10^{-3}$. The magnetic field values are contained in the fit parameter $b$, which is equal to $z \Blos$, where $z$ is the Zeeman splitting factor, and \Blos~is the line-of-sight (LOS) magnetic field strength (assuming, of course, that the signature in Stokes $V$ is due to the magnetic field in the region; see \S~\ref{ss:instru}). We used the AIPS task ZEMAN (\citealt{greisen2015}) to carry out the fit in equation~(\ref{e1}). This task allows multiple Gaussian components in $I$ to be fitted simultaneously to $V$, with each Gaussian component fitted for a different $b$, and hence a different LOS magnetic field strength. For the three Gaussian components of maser A+B from our 2009 observations (listed in Table~\ref{tGauss}, and shown in the upper panel of Figure~\ref{fig2}), the derivative profiles scaled by the respective values fitted for $b=z\Blos$ are each shown in the lower panel of Figure~\ref{fig2}; the composite profile obtained by adding together these three scaled derivative profiles is shown in the lower panel of Figure~\ref{fig3}. Of these three components for maser A+B from our 2009 observations, only component 2 showed a significant fit, with the fitted value given by $z\Blos = 152 \pm 12$~Hz. Following customary practice in the field of Zeeman observations, we consider fits to be significant only if the ratio of fitted value to the fitted error is at the 3-$\sigma$ level or greater. Meanwhile, for the three Gaussian components of maser B from our 2017 observations (listed in Table~\ref{tGauss}, and shown in the upper panel of Figure~\ref{fig4}), the derivative profiles scaled by the respective fitted values for \Blos, are each shown in the lower panel of Figure~\ref{fig4}, and the composite profile is shown in the lower panel of Figure~\ref{fig5}. Only component 2 of maser B from our 2017 observations showed a significant fit, with the fitted value given by $z\Blos = 149 \pm 19$~Hz. Since component 2 of maser A+B from our 2009 observations and component 2 of maser B from our 2017 observations are very likely the same maser spot (see \S~\ref{ss:spots}), this is a remarkable coincidence in \Blos\ over observations taken eight years apart.

\begin{deluxetable*}{cccccccccccc}
\tablenum{3}
\tablewidth{0pt}
\tablecaption{MAGNETIC FIELD VALUES
	\protect\label{tBlos}}
\tablehead{
& & & &  2009 observations & & 2017 observations \\ 
& & & &  ($z\Blos = 152 \pm 12$~Hz) & & ($z\Blos = 149 \pm 19$~Hz) \\ \hline
\colhead{$F_{\rm up}$\tablenotemark{a}} &
\colhead{$F_{\rm down}$\tablenotemark{a}} &
\colhead{$|z|$} & 
\colhead{} &
\colhead{\Blos} &
\colhead{} &
\colhead{\Blos} \\
\colhead{} &
\colhead{} &
\colhead{(Hz~mG$^{-1}$)}  &
\colhead{} &
\colhead{(mG)} &
\colhead{} &
\colhead{(mG)}}
\startdata
6 & 6 & 0.708 & &  $214 \pm 17$ & &  $211 \pm 27$ \\
4 & 4 & 0.873 & & 	$174 \pm 14$ & & 	$171 \pm 22$ \\
6 & 6 & 0.732 & & $207 \pm 17$ & &	$204 \pm 26$ \\
4 & 4 & 0.864 & & $175 \pm 14$ & &	$173 \pm 22$ \\
\enddata
\tablenotetext{a}{ The notation for hyperfines is explained in \citet{lankhaar2018}. Briefly, for the 25~GHz $5_2 - 5_1$~E Class I \meth\ masers, there are hyperfine states with $F=J$ and $F=J\pm1$, with four levels for the former, and two each for the latter. Since $J=5$ for the 25 GHz line, the possible values for $F_{\rm up}$ are 6, 5, 4, with four levels corresponding to $F_{\rm up}$=5, two levels corresponding to $F_{\rm up}$=6, and two levels for $F_{\rm up}$=4; likewise for $F_{\rm down}$. The strongest hyperfine transitions are for $\Delta F=0$. Thus, we have 8 hyperfine transitions, of which 4 are listed here. The other 4 hyperfine transitions are not listed in this table because the \Blos\ values calculated from them are unreasonably high (\S\ \ref{sec:res}).}
\end{deluxetable*}

Extracting the value of \Blos\ from the fit parameter $b$ in equation~(\ref{e1}) requires knowing the value of the Zeeman splitting factor $z$. This is not easy, since \meth\ maser lines may be comprised of one or more hyperfine transitions. \citet{lankhaar2018} derived the values of $z$ for a wide array of methanol maser lines by doing quantum mechanical calculations. For the 25 GHz $5_2 - 5_1$~E Class I \meth\ maser line that we observed, they list the values of $z$ for 8 hyperfine components. Four of these have $|z|$ between 0.7-0.9~Hz~mG$^{-1}$, and the corresponding values of \Blos\ are listed in Table~\ref{tBlos}; they range from 171~mG to 214~mG. This is 3-4 times higher than the largest \Blos\ we've measured at 44~GHz (\citealt{ms2017}), but not implausibly high (see the discussion in \S\ \ref{ss:Bn}). The other four hyperfines have $|z|$ values in the range 0.02-0.06~Hz~mG$^{-1}$; corresponding \Blos\ values would be 10-50 times higher than 171~mG, too high for the regions traced by these 25 GHz \meth\ masers; thus, we have not included them in Table~\ref{tBlos} (also see \S~\ref{ss:Bn}). In other words, we can rule out that the hyperfines with $|z|$ = 0.02-0.06 are responsible for the maser transition being observed. We note also that we have ignored the sign of $z$ in finding values for \Blos. Two of the four $z$ values listed in Table~\ref{tBlos} have negative signs in \citet{lankhaar2018}; by convention, a positive value for \Blos\ (when $z$ is positive) has meant that the LOS magnetic field is pointing away from the observer. Since $z$ has mixed signs in the treatment of \citet{lankhaar2018}, and it is not clear which hyperfine is masing to create the population inversion for the 25 GHz line, we would like to avoid any comment on the sign of \Blos. 

\section{Discussion}
\label{sec:disc}

The 25 GHz Class I \meth\ masers in \omc\ are the brightest masers of this species discovered to date. If the signature observed in Stokes $V$ is due to magnetic fields in \omc, then this would represent the first detection of the Zeeman effect in the 25 GHz \meth\ maser line. 

\subsection{Maser Spots}
\label{ss:spots}

\begin{figure}[htb!]
\plotone{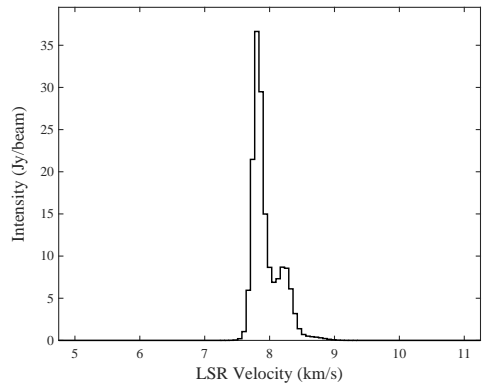}
\caption{Composite Stokes $I$ profile obtained by adding together the 5 fitted gaussian components (2 for maser A, 3 for maser B) listed in Table~\ref{tGauss} for our 2017 C-configuration observations, and scaling by the ratio of intensities as described in \S~\ref{ss:spots}. This composite spectrum is strikingly similar to the observed Stokes~$I$ profile from our 2009 D-configuration observations (see the upper panel of Figure~\ref{fig2}), even picking up the very low intensity extension past 8.5 km/s. \label{fig6}}
\end{figure}

We fitted the maser labeled as A+B in our 2009 VLA D-configuration observations with three Gaussian components in velocity (Table~\ref{tGauss}). This maser was resolved into two sources in our 2017 VLA C-configuration observations; we fitted the stronger of these (maser A) with two Gaussian components in velocity, and the other (maser B) with three components (Table~\ref{tGauss}). In fitting gaussian components to observed spectral line profiles, one acknowledges that the components themselves may not represent physical structures, but are the best decomposition of the observed spectral line. Still, it is worth checking if the five Gaussian components fitted to the 2017 C-configuration spectral lines (2 components for maser A and 3 for maser B) are consistent with the profile observed with the D-configuration of the VLA in 2009. In order to carry out this check, we added together all five profiles in velocity space and generated a composite profile. This composite profile was then scaled by the ratio of the intensity of component 1 of maser A+B from our 2009 observations to the intensity of component 1 of maser A from our 2017 observations (see Table~\ref{tGauss}). The result is shown in Figure~\ref{fig6}, and matches remarkably well with our 2009 D-configuration observations, for which the Stokes $I$ profile is shown in the upper panel of Figure~\ref{fig2} (and Figure~\ref{fig3}), including even the very low intensity wing at LSR velocities beyond 8.5 km/s. This increases the likelihood that the fitted velocity components correspond to actual maser spots in \omc. We have already stated above that component 1 of maser A+B from our 2009 observations is likely the same as component 1 of maser A in our 2017 observations. By comparing the $v_{\rm LSR}$ and FWHM velocity linewidth (Table~\ref{tGauss}), we conclude also that component 2 of maser A+B from our 2009 observations is likely the same as component~2 of maser~B in our 2017 observations. The identification of these masers spots from our 2009 and 2017 observations as being likely the same is important because they are the maser spots in which we have a significant Zeeman detection, assuming the signature in the Stokes~$V$ profile is due to magnetic fields in \omc.

\subsection{Instrumental Considerations and non-Zeeman Interpretation}
\label{ss:instru}

Structure in the Stokes $V$ profile can also be caused by instrumental effects and processes other than the Zeeman effect. A velocity gradient across an extended source could produce a Stokes $V$ profile due to the beam squint similar to that caused by the Zeeman effect. This appears unlikely in our current observations, since masers are point sources confined to a narrow velocity range. Another effect to consider is that in masers having high linear polarization to begin with, the linear polarization vector could rotate as the signal propagates along the line of sight due to changes in the orientation of the magnetic field. This would cause a circular polarization signature in the signal received at the telescope (\citealt{wiebe1998}). Although no measurements of the linear polarization are available for the 25~GHz \meth\ masers in \omc, we consider it unlikely that such a rotation of linear polarization is causing the Stokes $V$ profile in the present observations, since one would then expect the stronger masers listed in Table~\ref{tGauss} to also show significant circular polarization signatures. Yet another effect to consider is the generation of circular polarization by a rotation of the axis of symmetry for the molecular quantum states (\citealt{vlemmings2011}). This would occur if the maser stimulated emission rate $R$ were to become larger than $g\Omega$, the frequency shift due to the Zeeman effect. The stimulated emission rate $R$ is given by
\begin{equation} R \simeq \frac{AkT_b \Delta \Omega}{4\pi h \nu}  \label{e.2} \end{equation}
(\citealt{vlemmings2011}), where $A$ is the Einstein coefficient, $T_b$ is the maser brightness temperature, $\Delta \Omega$ is the maser beaming angle, and $\nu$ is the frequency of the observed maser transition. We use $A=5.570 \times 10^{-8}$~s$^{-1}$, the highest Einstein coefficient in \citet{lankhaar2018} for the $\nu~=~24959.079 \times 10^6$~Hz \meth\ maser transition, and $T_b = 10^6$-$10^7$~K, $\Delta \Omega \simeq 0.03-0.003$ for the 25 GHz maser line (\citealt{leurini+2016}). Using these values in equation~(\ref{e.2}), we obtain $R \le 3 \times 10^{-3}$~s$^{-1}$. Meanwhile, $g\Omega \approx 170$~s$^{-1}$ in our observations of \omc. This means that $R \ll g\Omega$, and thus it is unlikely that a rotation of the axis of symmetry for the molecular quantum states is responsible for the shape of the Stokes $V$ profile. Finally, \citet{houde2014} found that maser radiation scattering off foreground molecules can increase the antisymmetry in the Stokes $V$ spectral profile of SiO masers. Consequently, if the Stokes $V$ profile were ascribed to the Zeeman effect, one would obtain a much larger value for the magnetic field traced by these SiO masers. While we cannot rule out this effect in our 25~GHz \meth\ maser observations, our \Blos\ values are not orders of magnitude higher than the fields expected in such regions, unlike in SiO masers. We will discuss this in more detail in the next subsection.

\subsection{Magnetic Fields and Densities}
\label{ss:Bn}

If the observed Stokes $V$ profile is caused by a magnetic field in the source, and is not due to any instrumental effects or other non-Zeeman causes discussed in \S\ \ref{ss:instru}, then we have a magnetic field from our 25 GHz Class I \meth\ maser observation of OMC-1 that is 3-4 times larger (depending on which hyperfine is responsible for the maser) than the highest magnetic field we've detected to date in Class I methanol masers at 44 GHz (\citealt{ms2017}). Such a value is not implausible, however; if we assume that the fields are amplified in proportion to the density in the shocked regions where these masers occur, then
\begin{equation}
	 \frac{B_{\rm post}}{B_{\rm pre}} = \frac{n_{\rm post}}{n_{\rm pre}} \label{eBn} 
\end{equation}
where $B_{\rm pre}$ and $B_{\rm post}$ are the magnetic fields in the pre- and post-shock regions respectively, and $n_{\rm pre}$ and $n_{\rm post}$ 
are the densities in these regions. We can use equation~(\ref{eBn}) to calculate $B_{\rm pre}$ and compare it to magnetic fields calculated from other observations. Since our measured fields are in the range 171-214~mG depending on which hyperfine is responsible for the maser transition (Table~\ref{tBlos}), we will use \Bpost~=~171~mG in our calculations; this is convenient because \Bpre\ values corresponding to \Bpost~=~214~mG can then be found by multiplying our results by a factor of 1.25. \citet{leurini+2016} have found that bright Class I methanol masers likely occur in regions with densities in the range $10^{7-8}$~cm$^{-3}$, so we will calculate \Bpre\ corresponding to both $n_{\rm post} = 10^7$~cm$^{-3}$, and $10^8$~cm$^{-3}$. For densities in the pre-shocked gas we draw upon the work of \citet{kwan+1977}, who found that a minimum pre-shock density of $10^5$~cm$^{-3}$ is required to produce the observed intensities in the H$_2~\nu = 1-0$ emission lines in \omc. Thus, densities in the pre-shocked material of interest are likely of the order of $10^5$~cm$^{-3}$, or higher, but unlikely to be as high as, or higher than, $10^6$~cm$^{-3}$ (e.g., the references used below in discussing our calculated values for \Bpre). Therefore, we calculate \Bpre\ for a range of $n_{\rm pre}$ values from $1.0 \times 10^5$~cm$^{-3}$ to $1.0 \times 10^6$~cm$^{-3}$. The results of our calculations using equation~(\ref{eBn}) are given in Table~\ref{tPreB}; we find that for a lower post-shock density of $10^7$~cm$^{-3}$, the magnetic field in the pre-shock region could be as low as 1.7~mG or as high as 17~mG for the range of pre-shock densities used in our calculation. Meanwhile, if the post-shock density is as high as $10^8$~cm$^{-3}$, magnetic fields in the pre-shock region would be lower, and in the range of 0.17~mG to 1.7~mG. Indeed, \citet{chuss+2019} report magnetic field values of 0.931-1.013~mG in the BN/KL region of OMC-1, where $n=2.27 \times 10^5$~cm$^{-3}$. These magnetic field values were calculated by applying the Chandrasekhar-Fermi statistical method to their dust polarimetry data taken with the HAWC+ instrument on the Stratospheric Observatory for Infrared Astronomy (SOFIA). Meanwhile, \citet{pattle+2017} calculated a magnetic field of 6.6~mG in gas of density $8.3 \times 10^5$~cm$^{-3}$ toward \omc, based on polarization measurements as part of the B-fields in Star-forming Region Observations (BISTRO) survey with the James Clerk Maxwell Telescope (JCMT).

\begin{deluxetable}{ccccccccccc}
\tablenum{4}
\tablewidth{0pt}
\tablecaption{PRE-SHOCK MAGNETIC FIELDS
	\protect\label{tPreB}}
\tablehead{
\colhead{\phm{llll}$n_{\rm post}$\phm{llll}} & &
\colhead{\phm{llll}$n_{\rm pre}$\phm{llll}} & &
\colhead{\phm{llll}\Bpost\phm{llll}} &  &
\colhead{\phm{llll}\Bpre\phm{llll}} \\
\colhead{(cm$^{-3}$)} & &
\colhead{(cm$^{-3}$)} & &
\colhead{(mG)}  & &
\colhead{(mG)} }
\startdata
$10^7$ & &  $1.0 \times 10^5$  & & 171 & & 1.71 \\
$10^7$ & &  $2.5 \times 10^5$ & & 	171 & & 4.28 \\
$10^7$ & &  $5.0 \times 10^5$  & & 171	& & 8.55 \\
$10^7$ & &  $7.5 \times 10^5$  & & 171 & & 12.8 \\ 
$10^7$ & &  $1.0 \times 10^6$  & & 171 & & 17.1 \\ \hline
$10^8$ & &  $1.0 \times 10^5$  & & 171	& & 0.17 \\
$10^8$ & &  $2.5 \times 10^5$  & &	171 & & 0.43 \\
$10^8$ & &  $5.0 \times 10^5$  & & 171 & & 0.86 \\
$10^8$ & &  $7.5 \times 10^5$  & & 171 & & 1.28 \\
$10^8$ & &  $1.0 \times 10^6$  & & 171 & & 1.71
\enddata
\end{deluxetable}

With values for $n_{\rm pre}$ and $B_{\rm pre}$ available from \citet{chuss+2019} and \citet{pattle+2017} for the pre-shock regions of interest in \omc, we can also use our detected values of \Bpost\ in equation~(\ref{eBn}) to narrow the range of post-shock densities in which the 25 GHz \meth\ masers are being excited. The calculated values of $n_{\rm post}$ are given in Table~\ref{tPostN}. For pre-shocked regions where the densities are $2.27 \times 10^5$~cm$^{-3}$ and fields are $\sim$1~mG, we see that 25 GHz Class I \meth\ masers would be excited in regions of density $10^{7.6}$~cm$^{-3}$ if these regions had magnetic fields of 171~mG; $n_{\rm post}$ would be $10^{7.7}$~cm$^{-3}$ if the magnetic fields were as high as 214~mG. Meanwhile, if the fields and densities were higher in the pre-shock regions (6.6~mG and $8.3 \times 10^5$~cm$^{-3}$ respectively), then the masers would be excited in post-shock regions with densities $10^{7.3}$~cm$^{-3}$ if the magnetic field in these regions were 171~mG, or $10^{7.4}$~cm$^{-3}$ for post-shock magnetic fields as high as 214~mG. All of these values for $n_{\rm post}$ are in good agreement with densities of $10^{7-8}$~cm$^{-3}$\ \citep{leurini+2016} in the post-shock regions in which these 25~GHz Class~I \meth\ masers occur.

\begin{deluxetable}{ccccccccccccc}
\tablenum{5}
\tablewidth{0pt}
\tablecaption{POST-SHOCK DENSITIES
	\protect\label{tPostN}}
\tablehead{
\colhead{\phm{llll}\Bpre\phm{llll}} & &
\colhead{\phm{llll}$n_{\rm pre}$\phm{llll}} & &
\colhead{\phm{llll}\Bpost\phm{llll}} &  &
\colhead{\phm{llll}$n_{\rm post}$\phm{llll}} \\
\colhead{(mG)} & &
\colhead{(cm$^{-3}$)} & &
\colhead{(mG)}  & &
\colhead{(cm$^{-3}$)} }
\startdata
0.931\tablenotemark{a} & & $2.27 \times 10^5$ & & 171 & & $10^{7.6}$ \\
1.013\tablenotemark{a} & & $2.27 \times 10^5$ & & 171 & & $10^{7.6}$ \\
0.931\tablenotemark{a} & & $2.27 \times 10^5$ & & 214 & & $10^{7.7}$ \\
1.013\tablenotemark{a} & & $2.27 \times 10^5$ & & 214 & & $10^{7.7}$ \\ \hline
6.6\tablenotemark{b} & & $8.3 \times 10^5$ & & 171 & & $10^{7.3}$ \\
6.6\tablenotemark{b} & & $8.3 \times 10^5$ & & 214 & & $10^{7.4}$
\enddata
\tablenotetext{a}{ \Bpre\ and $n_{\rm pre}$ taken from \citet{chuss+2019}.}
\tablenotetext{b}{ \Bpre\ and $n_{\rm pre}$ taken from \citet{pattle+2017}.}
\end{deluxetable}

\subsection{Magnetic Field Values and Energetics}
\label{ss:Emag}

If we accept that the observed Stokes $V$ profile is due to the Zeeman effect, then we obtain that \Blos = 171-214~mG in the regions traced by these 25~GHz Class~I \meth\ masers. As we have noted already, these values are high, about 3-4 times higher than the largest field we have measured in the 44~GHz Class~I \meth\ maser line\ \citep{ms2017}. Yet, as we have demonstrated in \S~\ref{ss:Bn}, the \Blos\ values are not implausibly high. It is possible that these high values for \Blos\ could be a consequence of the \omc\ region being a special case. No other region to date is known to have 25 GHz \meth\ masers as bright as those in \omc. The outflow in Orion-KL is one of the most powerful ever discovered in a star forming region. In an interesting coincidence, \citet{troland+2016} found from Zeeman effect observations in the thermal lines of H I and OH that magnetic fields in the Orion Veil, a photon-dominated region (PDR) $\approx$2~pc \textit{in front of} the Trapezium stars in Orion, are 3-5 times stronger than they are in other regions with comparable values of density. \citet{troland+2016} speculate that the reason for this could lie in the history of \omc\ before star formation, in that \omc\ formed in a low turbulence environment due to its position outside the Galactic Plane, and less magnetic flux was removed from the developing cloud. 

A central purpose of Zeeman effect observations is to compare the magnetic energy to other relevant energy values in the region being observed. We can use our derived value of \Blos\ to find the magnetic energy density and compare it to the kinetic energy density. The magnetic energy density is given by $B^2/8\pi$, where $B^2 = 3\Blos^2$ (\citealt{crutcher1999}). If we use \Blos = 171~mG, we get that the magnetic energy density is equal to $3.5 \times 10^{-3}$~erg cm$^{-3}$. Meanwhile, the kinetic energy density is given by $(3/2)\, m n \sigma^2$, where $m$ is the mass and $\sigma$ is the velocity dispersion. This expression for the kinetic energy density includes the contribution of both thermal and turbulent motions. To find the mass, we use $m = 2.8\, m_p$, where $m_p$ is the proton mass; the numerical factor of 2.8 also accounts for the presence of 10\% He. The velocity dispersion is related to the FWHM velocity linewidth $\Delta v$ by $\sigma = \Delta v/(8 \ln\, 2)^{1/2}$. Using the observed FWHM velocity linewidths (0.13-0.53~\kmS; Table~\ref{tGauss}) of the 25 GHz \meth\ masers, however, will not yield a representative value of the kinetic energy density in the post-shock region, since masers may have a narrower linewidth than that in the region in which they are formed. We use $\Delta v = 7$~\kmS\ from observations of the quasi-thermal $10_1-9_2$~A$^{-}$ \meth\ transition at 23.4 GHz by \citet{wilson+1989}. This gives a value of $\sim 3.1 \times 10^{-5}$~erg~cm$^{-3}$ for the kinetic energy density in \omc, implying the magnetic energy density in the post-shocked gas dominates over the kinetic energy density. A better indicator of the significance of the magnetic field, however, might be to compare it to the pressure in the shock, given by $(1/2)\, \rho v^2$, where $v$ is the shock velocity. Using a shock velocity of 30~\kmS~(\citealt{leurini+2016}), we get a pressure of $5.3 \times 10^{-4}$~erg cm$^{-3}$; the pressure would have a higher value of $1.1 \times 10^{-3}$~erg cm$^{-3}$ if the postshock density were as high as $10^{7.7}$~cm$^{-3}$. Thus the magnetic energy is of the same order as, or marginally larger than, the pressure in the shock. Therefore, we would expect the magnetic field to play a significant role in shaping the dynamics in these post-shocked regions of \omc. 

\section{Conclusions}
\label{sec:conc}

We have discovered a signature in the Stokes $V$ profile of a 25~GHz Class~I \meth\ maser toward the high mass star forming region \omc. If this feature in Stokes $V$ is caused by a magnetic field in the source and is not due to instrumental effects or other non-Zeeman contributions, then these observations would be the first detection and discovery of the Zeeman effect in the 25~GHz Class~I \meth\ maser line. We find that $z\Blos = 152 \pm 12$~Hz from our first epoch of observations in 2009 in the D-configuration of the VLA, and $z\Blos = 149 \pm 19$~Hz from our second epoch of observations in 2017 in the C-configuration of the VLA, very likely for the same maser spot. The agreement in magnetic field strengths over observations taken 8 yr apart with different angular resolutions is remarkable. Based on which one of four hyperfines is masing, these values correspond to magnetic fields in the range 171-214~mG; there are four other hyperfines that we exclude because the \Blos\ yielded by them would be unreasonably high for the regions in which these masers occur. Our detected values of \Blos\ are nevertheless high, about 3-4 times higher than the strongest field we have measured in 44 GHz Class~I \meth\ masers. Yet, they are not implausible. For a density of $10^7$~cm$^{-3}$ in the post-shock regions in which the 25~GHz Class~I \meth\ masers occur, \Blos=171~mG corresponds to a pre-shock field of 1.7~mG in a region of density $1.0 \times10^5$~cm$^{-3}$. Even if densities were higher, $1.0 \times 10^6$~cm$^{-3}$ for the pre-shock region and $10^8$~cm$^{-3}$ for the post-shock region respectively, the magnetic field in the pre-shock region would still be 1.7~mG. Conversely, if fields and densities in pre-shock regions were 0.9-1.0~mG and $2.27 \times 10^5$~cm$^{-3}$ respectively, as observed by \citet{chuss+2019}, then a post-shock field of 171~mG would imply a region of density $10^{7.6}$~cm$^{-3}$. It is in regions with densities $10^{7-8}$~cm$^{-3}$ that 25~GHz Class~I \meth\ masers are believed to occur. Thus, the higher \Blos\ values we have measured may be due to the \omc\ region being special; this includes \omc\ having the brightest known 25~GHz Class~I \meth\ masers. Finally, our detected values of \Blos\ indicate that the magnetic energy will dominate over the kinetic energy, even if the FWHM linewidth in these regions is as broad as 7~\kmS. Also, the magnetic energy is at least of the same order, and perhaps slightly higher in value, than the pressure in the shock. Therefore, the magnetic field is likely important in shaping the dynamics in the post-shocked regions where these masers occur.

\acknowledgments

We would like to thank an anonymous referee for comments that have helped in improving the final manuscript. 

\facility{VLA}.


\begin{thebibliography}{}
\bibitem[Bally et al.(1987)]{bally+1987} Bally, J., Langer, W.~D., Stark, A.~A., et al.\ 1987, \apj, 312, L45
\bibitem[Bally(2008)]{bally2008} Bally, J.\ 2008, Handbook of Star Forming Regions, Volume I, 459
\bibitem[Barrett et al.(1971)]{barrett+1971} Barrett, A.~H., Schwartz, P.~R., \& Waters, J.~W.\ 1971, \apj, 168, L101
\bibitem[Becklin \& Neugebauer(1967)]{becklin+1967} Becklin, E.~E., \& Neugebauer, G.\ 1967, \apj, 147, 799
\bibitem[Beckwith et al.(1978)]{beckwith+1978} Beckwith, S., Persson, S.~E., Neugebauer, G., et al.\ 1978, \apj, 223, 464
\bibitem[Chuss et al.(2019)]{chuss+2019} Chuss, D.~T., Andersson, B.-G., Bally, J., et al.\ 2019, ApJ, 872, 187
\bibitem[Crutcher(1999)]{crutcher1999} Crutcher, R.~M.\ 1999, \apj, 520, 706 
\bibitem[Crutcher(2012)]{crutcher2012} Crutcher, R.~M.\ 2012, \araa, 50, 29
\bibitem[Erickson et al.(1982)]{erickson+1982} Erickson, N.~R., Goldsmith, P.~F., Snell, R.~L., et al.\ 1982, \apjl, 261, L103
\bibitem[Genzel \& Stutzki(1989)]{genzel+1989} Genzel, R., \& Stutzki, J.\ 1989, Annual Review of Astronomy and Astrophysics, 27, 41
\bibitem[Greisen(2003)]{greisen+2003} Greisen, E.~W.\ 2003, Information Handling in Astronomy - Historical Vistas, 285, 109 
\bibitem[Greisen(2015)]{greisen2015} Greisen, E.~W.\ 2015, AIPS Memo, 118 \url{http://www.aips.nrao.edu/aipsmemo.html}
\bibitem[Hacar et al.(2018)]{hacar+2018} Hacar, A., Tafalla, M., Forbrich, J., et al.\ 2018, \aap, 610, A77
\bibitem[Houde(2014)]{houde2014} Houde, M.\ 2014, \apj, 795, 27 
\bibitem[Johnston et al.(1992)]{johnston+1992} Johnston, K.~J., Gaume, R., Stolovy, S., et al.\ 1992, \apj, 385, 232
\bibitem[Johnstone \& Bally(1999)]{johnstone+1999} Johnstone, D., \& Bally, J.\ 1999, \apjl, 510, L49
\bibitem[Kleinmann, \& Low(1967)]{kleinmann+1967} Kleinmann, D.~E., \& Low, F.~J.\ 1967, \apj, 149, L1
\bibitem[Kounkel et al.(2017)]{kounkel+2017} Kounkel, M., Hartmann, L., Loinard, L., et al.\ 2017, \apj, 834, 142
\bibitem[Krumholz \& Federrath(2019)]{krumholz+2019} Krumholz, M.~R., \& Federrath, C.\ 2019, Frontiers in Astronomy and Space Sciences, 6, 7
\bibitem[Kwan et al.(1977)]{kwan+1977} Kwan, J.~H., Gatley, I., Merrill, K.~M., et al.\ 1977, \apj, 216, 713
\bibitem[Lankhaar et al.(2018)]{lankhaar2018} Lankhaar, B., Vlemmings, W., Surcis, G., et al.\ 2018, Nature Astronomy, 2, 145 
\bibitem[Leurini et al.(2016)Leurini, Menten, \& Walmsley]{leurini+2016} Leurini, S., Menten, K.~M., \& Walmsley, C.~M.\ 2016, \aap, 592, A31
\bibitem[Menten(1993)]{menten1993} Menten, K.~M.\ 1993, Astrophysical Masers; Proceedings of the Conference, Arlington, VA, Mar. 9-11, 1992, 199
\bibitem[Momjian \& Sarma(2017)]{ms2017} Momjian, E., \& Sarma, A.~P.\ 2017, \apj, 834, 168
\bibitem[Momjian \& Sarma(2019)]{momjian+2019} Momjian, E., \& Sarma, A.~P.\ 2019, ApJ, 872, 12 
\bibitem[Motte et al.(2018)]{motte+2018} Motte, F., Bontemps, S., \& Louvet, F.\ 2018, \araa, 56, 41
\bibitem[Pabst et al.(2019)]{pabst+2019} Pabst, C., Higgins, R., Goicoechea, J.~R., et al.\ 2019, \nat, 565, 618
\bibitem[Pattle et al.(2017)]{pattle+2017} Pattle, K., Ward-Thompson, D., Berry, D., et al.\ 2017, \apj, 846, 122
\bibitem[Sarma \& Momjian(2009)]{sarma+2009} Sarma, A.~P., \& Momjian, E.\ 2009, \apjl, 705, L176
\bibitem[Sault et al.(1990)]{sault1990} Sault, R.~J., Killeen, N.~E.~B., Zmuidzinas, J., \& Loushin, R.\ 1990, \apjs, 74, 437 
\bibitem[Troland \& Heiles(1982)]{th1982} Troland, T.~H., \& Heiles, C.\ 1982, \apj, 252, 179 
\bibitem[Troland \& Crutcher(2008)]{troland+2008} Troland, T.~H., \& Crutcher, R.~M.\ 2008, \apj, 680, 457
\bibitem[Troland et al.(2016)]{troland+2016} Troland, T.~H., Goss, W.~M., Brogan, C.~L., et al.\ 2016, \apj, 825, 2
\bibitem[Vlemmings et al.(2011)]{vlemmings2011} Vlemmings, W.~H.~T., Torres, R.~M., \& Dodson, R.\ 2011, \aap, 529, A95 
\bibitem[Wiebe \& Watson(1998)]{wiebe1998} Wiebe, D.~S., \& Watson, W.~D.\ 1998, \apjl, 503, L71 
\bibitem[Wilson et al.(1989)]{wilson+1989} Wilson, T.~L., Johnston, K.~J., Henkel, C., et al.\ 1989, \aap, 214, 321
\end{thebibliography}
\end{document}